\documentclass[sigconf, nonacm]{acmart}
\usepackage{float}
\usepackage[bottom]{footmisc}

\begin{document}
\title{A computational analysis on the relationship between melodic originality and thematic fame in classical music from the Romantic period.}

\author{Hudson Griffith}
\email{griffithh@ufl.edu}

\maketitle

\section{Introduction}

Throughout modern society, many individuals have attempted to predict popularity. Whether it be the popularity of a company, a new product, or a tweet. This is often done by finding a correlation between popularity and another variable. In some cases, it may be analyzing media type to predict the popularity of an advertisement \cite{yu-2011} and in other cases, it is analyzing melodic originality to predict the popularity of classical music \cite{simonton-1980A}. This paper will be looking into the relationship between melodic originality and popularity in classical music. For this research, I would like to propose a novel method for calculating melodic originality. This novel method will be used in order to investigate the research question: To what extent does melodic originality affect thematic fame in classical music from the Romantic Period?

The research question proposed above will be answered through a computer content analysis of 428 classical pieces from the Romantic period. More specifically, this analysis was performed in Python \cite{python-software-foundation-2021} and R \cite{r-core-team-2017} by code I created specifically for this analysis. For this analysis, many of the definitions and methodology are based or heavily derived from Dean Keith Simonton’s research on melodic originality in the 1980s and 1990s. For this analysis melodic originality will be defined as the occurrence of uncommon two-note transitions. This is very similar to Simonton’s definition which is that melodic originality is “the occurrence both of chromatic notes or less commonly used diatonic notes and of rare intervals between consecutive notes” \cite{simonton-1980A}. Despite the similar definition, this analysis will be using a novel and different formula than Simonton so it is expected for the results of this paper to differ on some level from the conclusions stated in Simonton’s paper. This analysis will use Simonton’s definition of thematic fame which he defines as “the frequency that the theme would be heard in performance or recording (i.e., the level of appreciation by concert goers, record buyers, and performing musicians)”\cite{simonton-1980A}. These two variables will be compared to test their relationship and specifically how well melodic originality explains the variance of popularity. The novel method for calculating melodic originality represented in this paper addresses a gap in the research of the well-known researcher of melodic originality, Dean Keith Simonton. To calculate originality he only looked at the first 6 notes in each piece. The benefit of this new method proposed in this paper is that it provides much more information and data about each piece compared to Simonton’s method; however, further research is required before the robustness of this novel method is confirmed.

\section{Literature Review}

In this literature review key research in the field of computational musical analysis, melodic originality, and thematic fame will be discussed.

\subsection{Field Experts}

A key expert in the field of melodic originality in classical music is Dean Keith Simonton, a distinguished professor of Psychology at UC Davis. He defines much of the literature and methodology to this day for analyzing melodic originality in a wide variety of music. His 1980 paper “Thematic fame and melodic originality in classical music: A multivariate computer-content analysis” was the starting point for analyses into melodic originality of classical music and the first paper of its kind \cite{simonton-1980A}. Simonton has since published more papers relating to computational analysis of classical music up to 1994 and his paper “Computer Content Analysis of Melodic Structure: Classical Composers and Their Compositions” \cite{simonton-1994}. His two research papers published in 1980 heavily focus on melodic originality so my analysis will be more focused on these papers.
Research published in recent years continues to mention Simonton and his formulas. An example of this is Richard Hass who explored the relationship between melodic originality and fame in 20th-century American popular music \cite{hass-2015}. In Hass’s paper, he used an almost identical comparison for melodic originality to Simonton, showing the wide impact of Simonton's measure for melodic originality; furthermore, according to Google Scholar his second paper published in 1980 was cited by 207 other papers, once again confirming the importance of his research.
When it comes to the broader field of computational analyses of classical music there are other researchers influencing the field. One being Christof Weiß who works at the International Audio Laboratories Erlangen in Germany. He has published several papers on computational analyses in classical music, which unlike Simonton do not focus on melodic originality or thematic fame. Weiß’s original doctoral dissertation was on the topic of “Computational Methods for Tonality-Based Style Analysis” \cite{wei-2017} and he has continued research in this field specifically looking at tonal complexity and tonality-based style analysis \cite{weiss-2015}. He and his colleges make up a majority of recent publications in classical music computational analyses \cite{wei-2018}.
\subsection{Researcher Assumptions}

Most researchers seem to be making individual assumptions on what they consider “fame” or “success”. Simonton’s papers and papers citing him often use his definitions of fame and melodic originality, but many papers have different wording. They often assume that the relationship between melodic originality and thematic fame is a curvilinear relationship or inverted-U as Simonton showed in his papers. Despite Simonton’s results some papers such as Hass’s 2015 paper show results that contrast this. With his analysis of 20th-century popular music Hass stated that the “relationship between fame and originality is not linear” \cite{hass-2015} when referring to his figure. This raises questions on whether or not the relationship between melodic originality and thematic fame Simonton discovered only applies to the classical period he analyzed. With my research topic, I plan to analyze a more specific period of classical music which should answer this question.
\subsection{Popular Methodologies}
When it comes to the types of studies done in this field, all papers utilize or focus on computational content-analyses. The main methodology in this field is running the transcribed notes through algorithms to find the coefficients for variables such as melodic originality, aesthetic success, and originality variation. More recent papers such as the work done by Christof Weiß \cite{wei-2018} and Lesley Mearns \cite{mearns-2013} utilize MIDI files, which had yet to be invented during Simonton’s research. While MIDI does have it downsides like the ones mentioned in "The Dysfunctions of MIDI" \cite{moore-1988}, it is still widely used in the field of music and very useful for modern computational analyses.
However, there is an alternative to analyzing notated music files, which is signal analysis. Signal analysis, or signal processing as it is also known, is the analysis of the actual audio of a song recording. According to research by Meinard Müller, a professor for Semantic Audio Processing at the International Audio Laboratories Erlangen in Germany, music-based signal analysis is a growing field thanks to advances in speech signal processing; however, it is often more complex and difficult than the analyzing notated files \cite{muller-2011}. Signal analysis has been used to attempt to predict musical popularity in research by Junghyuk Lee \cite{lee-2018} as well as the emotion created by a musical piece in research done by Yazhong Feng \cite{yazhong-feng-2003}; furthermore, signal analysis has yet to be used to detect melodic originality. This is most likely because the popular method used to calculate melodic originality requires the specific notes of a piece which is often difficult for signal analysis to detect perfectly.
When it comes to programming languages, many papers do not state the specific language used; however, one paper by Hass \cite{hass-2015} states that he used R and even provides information on where to find the actual code he used. His code provided is an important reference when confirming the validity and testing the code I created for this analysis. Many sources do provide the specific formulas used to calculate the different variables which again helps confirm my methodology's validity. 
\subsection{Common Findings and Conclusions}
The most important finding in the field of analyzing melodic originality and thematic fame is that they have a curvilinear relationship. 
\begin{figure}[H]
  \centering
  \includegraphics[scale=0.6]{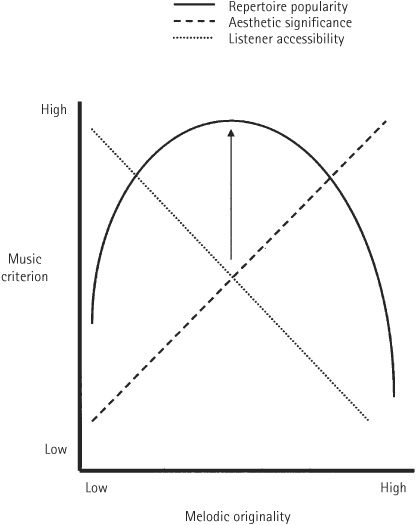}
  \caption{Relationship between melodic originality and objective repertoire popularity \cite{unknown-author-2010}.}
  \label{fig:curve}
\end{figure}

\autoref{fig:curve} shows this relationship and describes "how a theme’s melodic originality (as computed from two-note transition probabilities) is associated with three musical criteria: objective repertoire popularity (a composite measure of the performance and recording frequency of the composition which contains the theme) and subjective assessments of its aesthetic significance and listener accessibility" \cite{unknown-author-2010}.

This was discovered by Simonton in 1980 and is the most commonly cited conclusion when it comes to the relationship between melodic originality and popularity; however, there is research that does not match this conclusion. An example of this would be the previously stated research by Richard Hass and his finding that the “relationship between fame and originality is not linear” \cite{hass-2015}. When it comes to variables besides popularity, popular findings are that melodic originality increases with age and is greater for music composed during periods of biographical stress \cite{simonton-1986}.

\subsection{Gap in Research}

The gap in research that will be investigated in this paper is a new approach to calculating melodic originality. Below is the formula Simonton used to calculate melodic originality \cite{simonton-1994}.
\[Originality = 1 - (\frac{\sum_{1<i<5} P_{i}}{5})\]
This can be simplified as the sum of all two-note transition probabilities for the first 5 bigrams, P(i), divided by 5. This gives the average two-note transition probability over the first 6 notes. That is then subtracted from 1 to give the improbability over the first 6 notes or how unlikely the first 6 notes are in a musical piece. There is a gap in this because it only analyzes the first 6 notes. The novel method introduced in this paper analyzes all notes in a piece providing substantially more data. Another gap mentioned specifically by Simonton is that his process only analyzes two-note transitions. Simonton in 1980, states “it might be preferable to measure more than two-note transitions, and include three-, four-, five-, and six-note sequences” \cite{simonton-1980A}. If further research is done in this field it is recommended that the researchers look into a greater length of note sequences. The difficulty behind analyzing longer note sequences is the amount of data required. If 8-note sequences were analyzed instead, there would be over 19 million possible permutations. This is beyond the scope of this analysis but still quite possible with modern technology.

\subsection{Selecting Composers}
Before analyzing any music it was essential to first choose individual composers to analyze. This research specifically focuses on the Romantic period, so six composers from the Romantic period have been chosen. A total of six were selected in order to keep the amount of data more manageable. The time constraints of this research did not allow larger numbers of composers such as the 10 composers analyzed in Simonton's first 1980 paper \cite{simonton-1980A}. The six composers chosen were some of the most popular composers in the Romantic period. The composers chosen were as follows: Ludwig van Beethoven, Johannes Brahms, Frédéric Chopin, Franz Liszt, Franz Schubert, and Robert Schumann.
For this research, the Romantic period was chosen to limit the scope of this research into something manageable for the time constraints. The definition this paper will be using for the Romantic period is from The Editors of Encyclopaedia Britannica, “Romanticism, attitude or intellectual orientation that characterized many works of literature, painting, music, architecture, criticism, and historiography in Western civilization over a period from the late 18th to the mid-19th century” \cite{the-editors-of-encyclopaedia-britannica-no-date}.

\section{Methodology}
A total of 428 pieces were analyzed in this research process. To acquire the pieces in a format that can be computationally analyzed, I purchased access to “Kunst Der Fuge”, a database of classical music in MIDI format \cite{onclassical-2020}. Multiple pieces were downloaded from each of the composers. The number of pieces for each composer is as follows: Ludwig van Beethoven (n=16), Johannes Brahms (n=94), Frédéric Chopin (n=120), Franz Liszt (n=26), Franz Schubert (n=117), Robert Schumann (n=55). After downloading all of the pieces, a datasheet was created to organize the pieces. This data sheet included File Name, Piece Title, Composer, and empty columns for Melodic Originality and Popularity. Due to the way “Kunst Der Fuge” is formatted, the Piece Title for each piece had to be manually labeled. This greatly lengthened the process of data formatting, so it is recommended that a different database is used for further research. After creating the datasheet the pieces were ready to be analyzed. This processing and analysis was done primarily in the programming language Python; however, R was also used for post-processing of the data from Python. Multiple Python packages were used in this analysis, the key packages were pandas, NumPy, SciPy, Matplotlib, and music21. Pandas was used for data manipulation and analysis, NumPy was used for matrix and array creation, SciPy was used to conduct tests and create intervals, Matplotlib was used for plotting data, and Music21, “A Toolkit for Computational Musicology” \cite{mit-2021}, was the library used to iterate through each MIDI file and separate the notes. All code is publicly available at the GitHub referenced \cite{HudsonGri}.

\subsection{Calculating Transition Probabilities}
The first step in calculating melodic originality is finding the probability for note transitions among the data set. This was done by iterating through each piece to count each possible two-note transition. This was done for all 428 pieces and can be viewed as \autoref{tab:totals}.

\autoref{tab:totals} represents the total count of all two-note transitions for all pieces. This table is read row to column, meaning that a transition such as G\# to C occurred 16,340 times.

\begin{figure}[H]
  \centering
  \includegraphics[width=\linewidth]{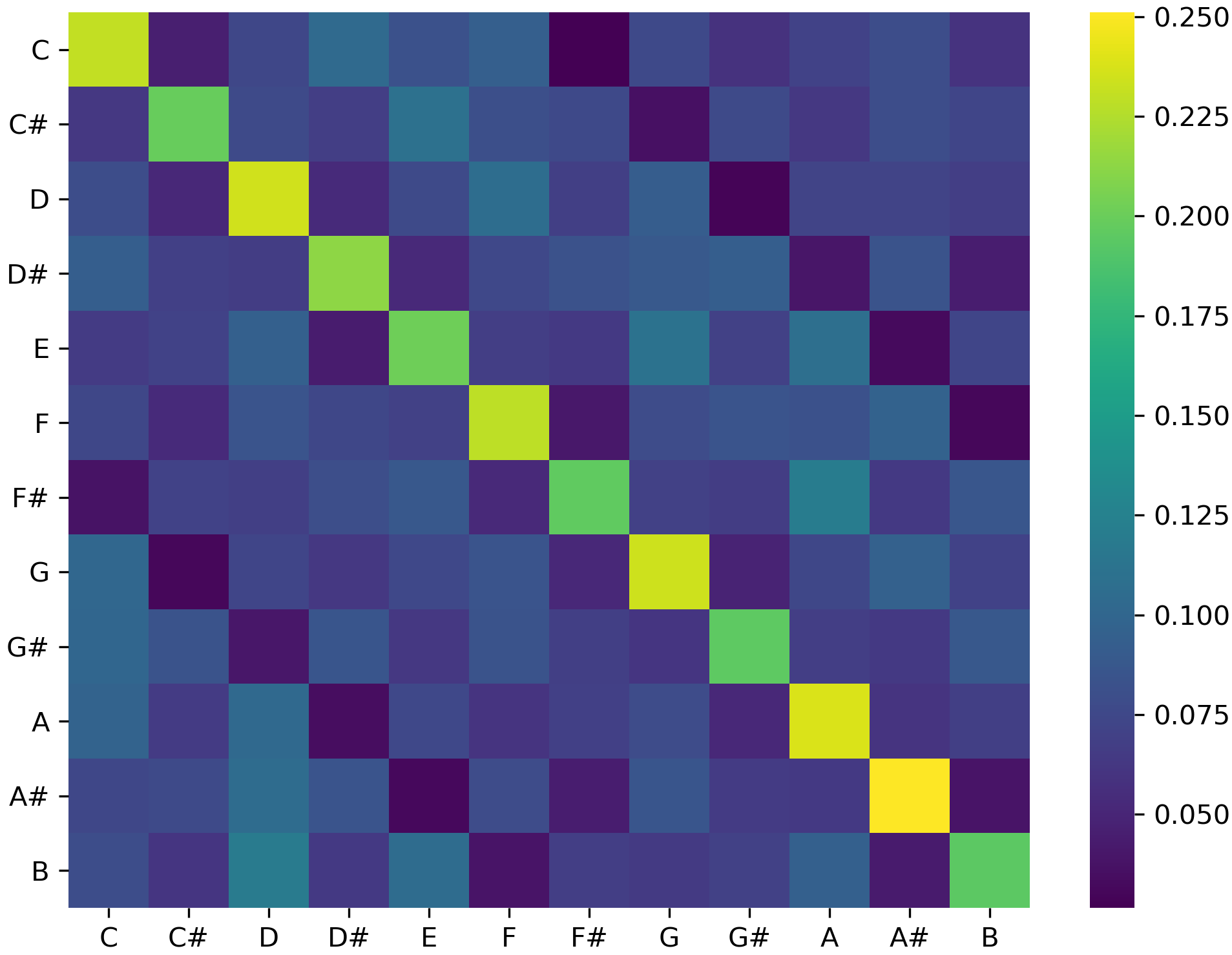}
  \caption{Heat map representing the relative probabilities of all two-note transitions.}
  \label{fig:matrix}
\end{figure}

Once all of the transitions are counted they must then be converted into probabilities. This is done by dividing each value by its row total. The result is a stochastic matrix representing the probability of a note-transition given the first note. The resulting matrix is displayed in \autoref{fig:matrix}. This figure is also read row to column. These probabilities represent probable and improbable transitions in the 428 pieces analyzed. For example, A\# to A\# was the most probable transition with a probability of 0.251078. \autoref{tab:prob} represents the data behind \autoref{fig:matrix}. These probabilities now allow the computation of melodic originality.

\subsection{Calculating Melodic Originality}
For this analysis, a novel formula for melodic originality will be used. In Simonton’s research, he defined melodic originality as the sum of the two-note transition probabilities for the first 6 notes in a piece. He then divided that by 5 to get the average probability across the first 5 two-note transitions. That was then subtracted from 1 to get the improbability or originality of the piece. The formula proposed in this research paper is below
\[Originality = 1 - (\frac{\sum_{1<i<n} P_{i}}{n-1})\]
This formula is similar to Simonton’s; however, it accounts for every single note in a piece and not only the first six. This provides substantially more information on the piece and represents a significant difference between Simonton’s formula. This novel formula can be interpreted as the sum of all two-note transition probabilities, P(i), over the total number of notes, (n), subtracted by 1. This is then subtracted by 1. Using this formula, the melodic originality of each piece is saved into a .csv file and the code is then ready to calculate popularity.

\subsection{Calculating Popularity}
To calculate the thematic fame or popularity of a piece, each full-length title must be known. To calculate popularity, the Python \cite{python-software-foundation-2021} code I developed for this analysis searches the title of each piece on Google and returns the number of search results. This provides an accurate metric of how many times the work is recorded, performed, or referenced. Google was chosen for this because according to a 2009 study by Asli Uyar, a computational scientist at The Jackson Laboratory, “Google provided the most accurate estimates for document counting for both single and multiple-term queries” \cite{uyar-2009}. This reaffirms the integrity of this metric. Once the popularity has been calculated for each piece, it can then be compared and tested against melodic originality.

\section{Results}
\begin{figure}[H]
  \centering
  \includegraphics[width=\linewidth]{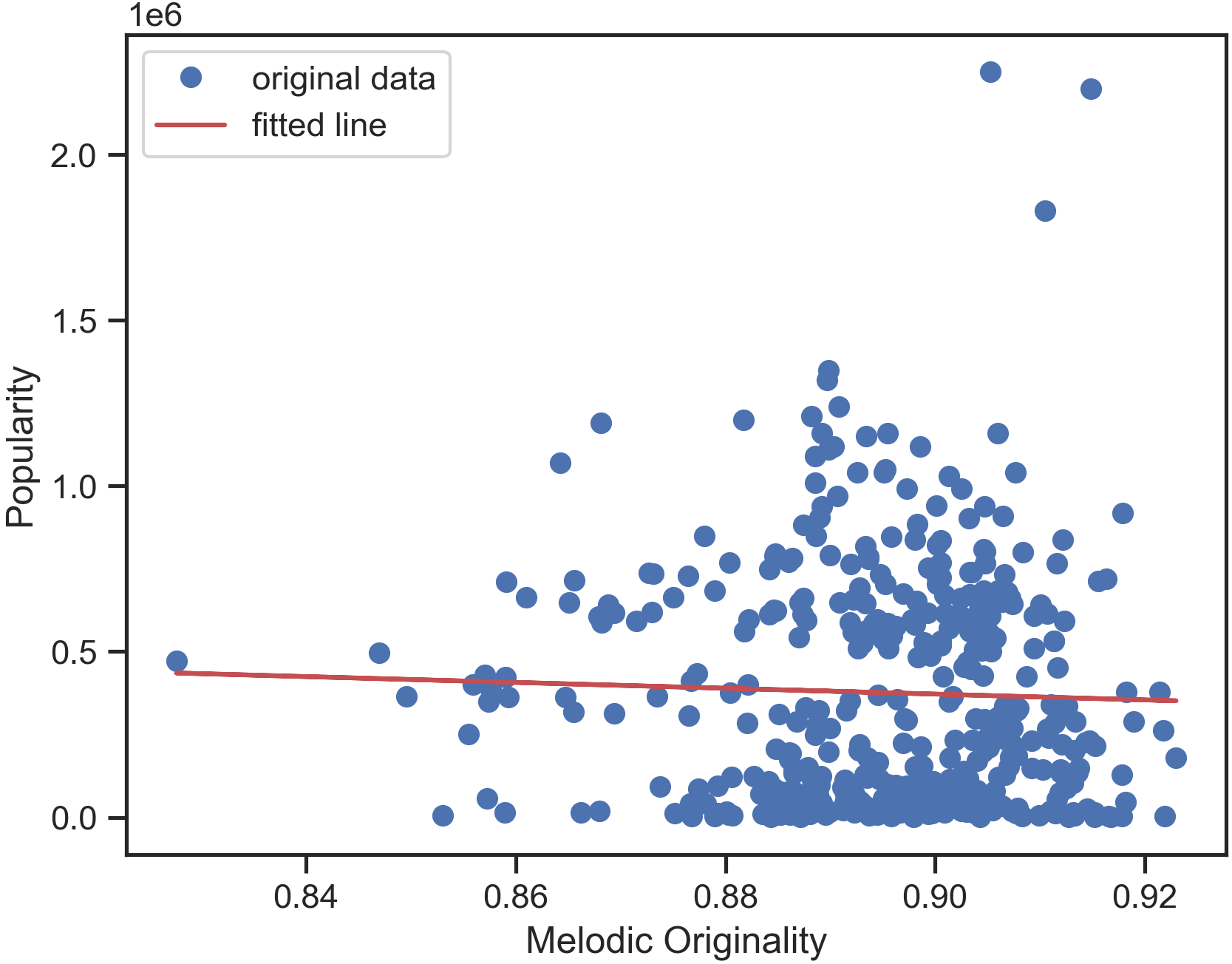}
  \caption{Linear regression on a scatter plot of melodic originality and popularity.}
  \label{fig:reg}
\end{figure}

The first calculation done to analyze the data obtained was a linear regression test. This was calculated using the Python library ‘scipy’. The results are shown in \autoref{fig:reg}.

This test resulted in a very low $R^{2}$ value. This can be interpreted as, 0.117\% of the data fits a linear regression model. From this information, it can be concluded that melodic originality explains very little of the variation in popularity. The r-value for this test also shows that there is a very weak correlation with a negative value of -0.0342607417.
As visualized in the graph the data very clearly does not fit the linear regression line and shows no clear correlation between variables. From \autoref{fig:reg} we can conclude that this data does not have a curvilinear relationship or inverted-U relationship that Simonton showed in his research. These results, however, are very similar to Hass’s results on his analysis between melodic originality and thematic fame in 20th-century popular music. The only key difference between Hass’s results was that his distribution was right-skewed unlike the left-skewed distribution in \autoref{fig:reg}.

\begin{figure}[H]
  \centering
  \includegraphics[width=\linewidth]{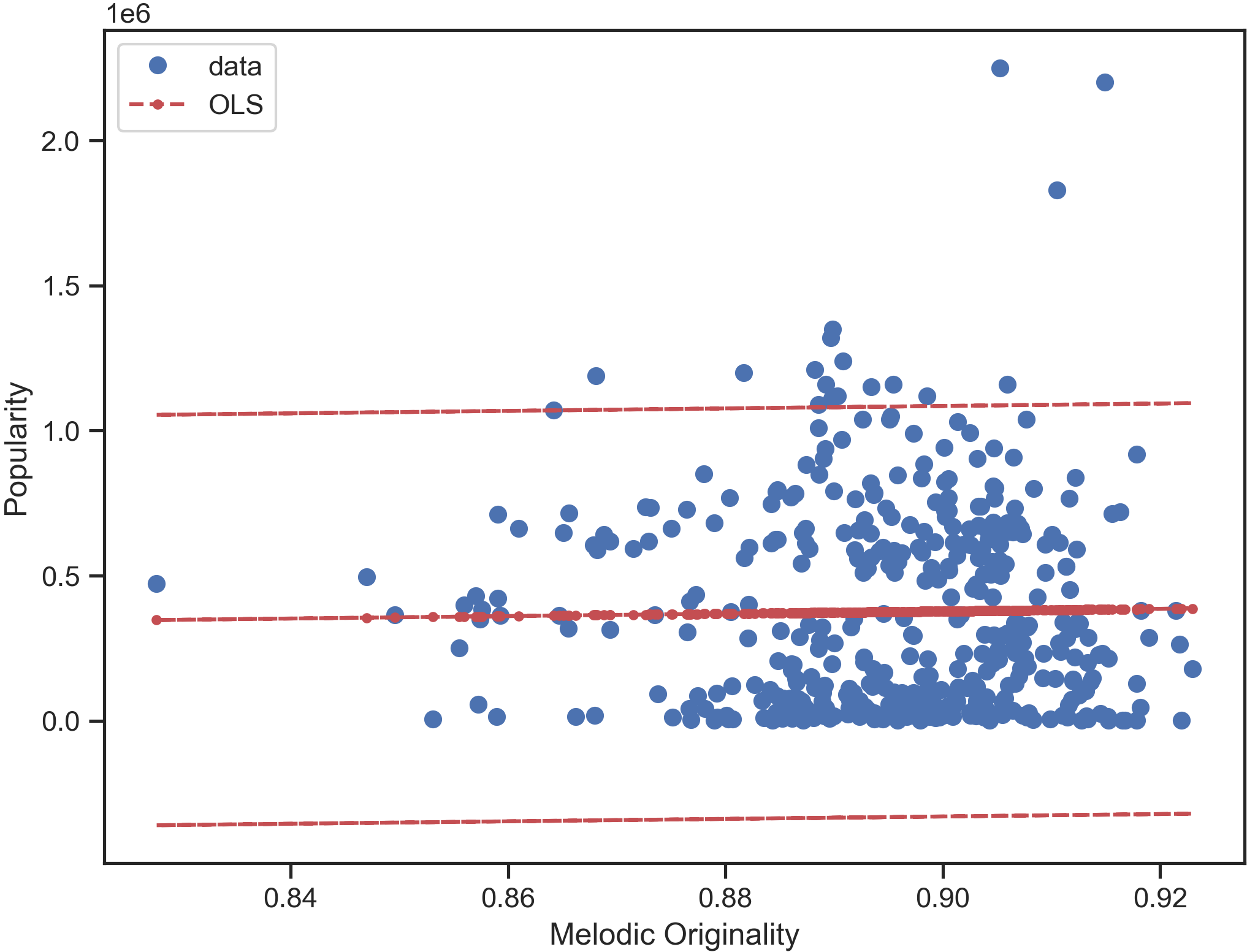}
  \caption{OLS Fitted Values Plot.}
  \label{fig:OLS}
\end{figure}

To further analyze this data, ordinary least squares regression was performed. This was calculated in Python using the ‘statsmodels’ library. The results of this are shown in \autoref{tab:OLS} and \autoref{fig:OLS}. The $R^{2}$ value calculated in this regression was 0.523. Unlike the previous linear regression, OLS resulted in a $R^{2}$ much closer to Simonton’s value of 0.48 \cite{simonton-1980A} and a p-value of under 0.01, the same as Simonton. This shows that while the correlation between melodic originality and popularity may not be clearly linear, there is some association that can be explained by the regression line.

\begin{figure}[H]
  \centering
  \includegraphics[width=\linewidth]{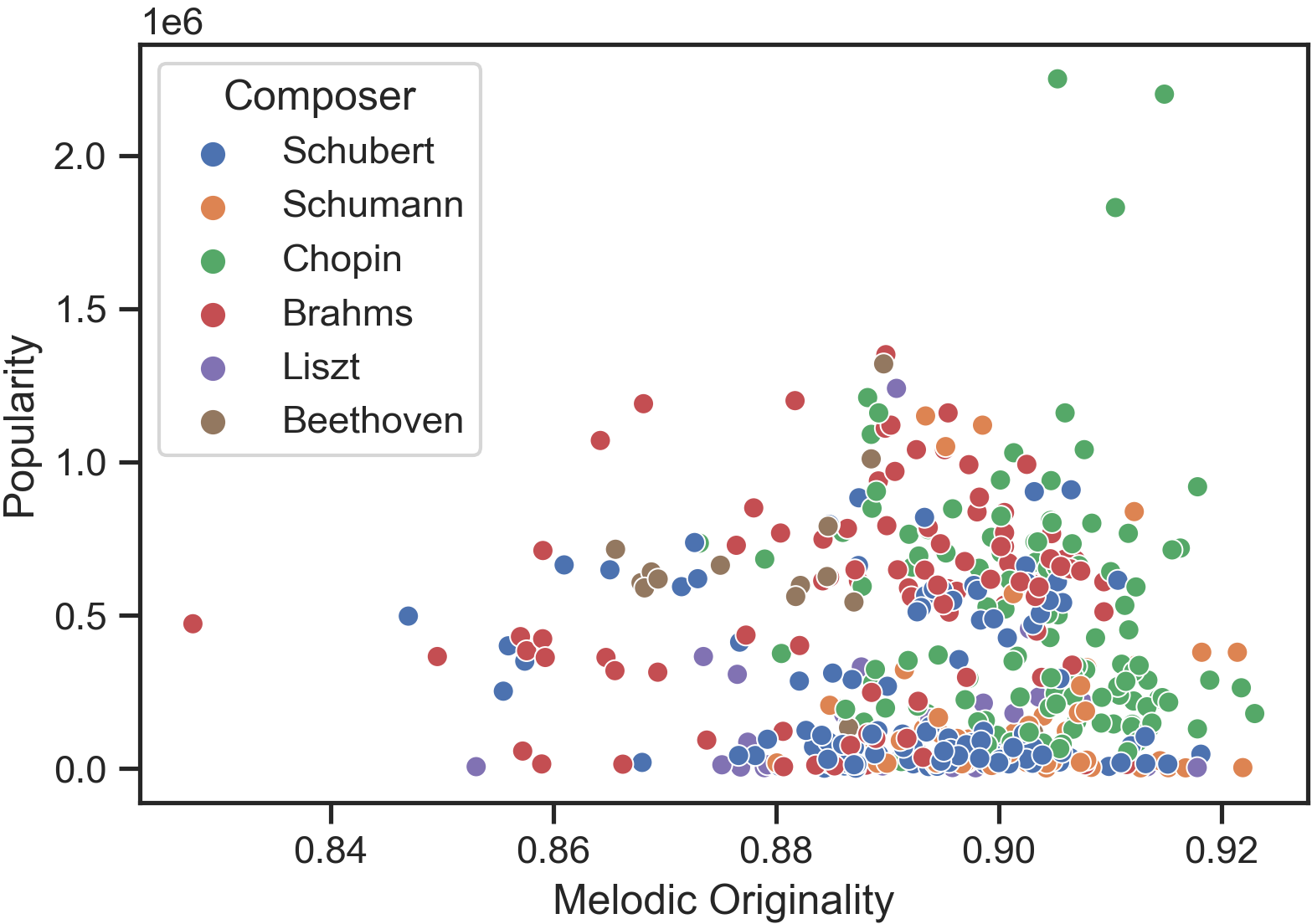}
  \caption{Scatter plot with points color coded by composer.}
  \label{fig:scatter}
\end{figure}

\begin{figure}[H]
  \centering
  \includegraphics[width=\linewidth]{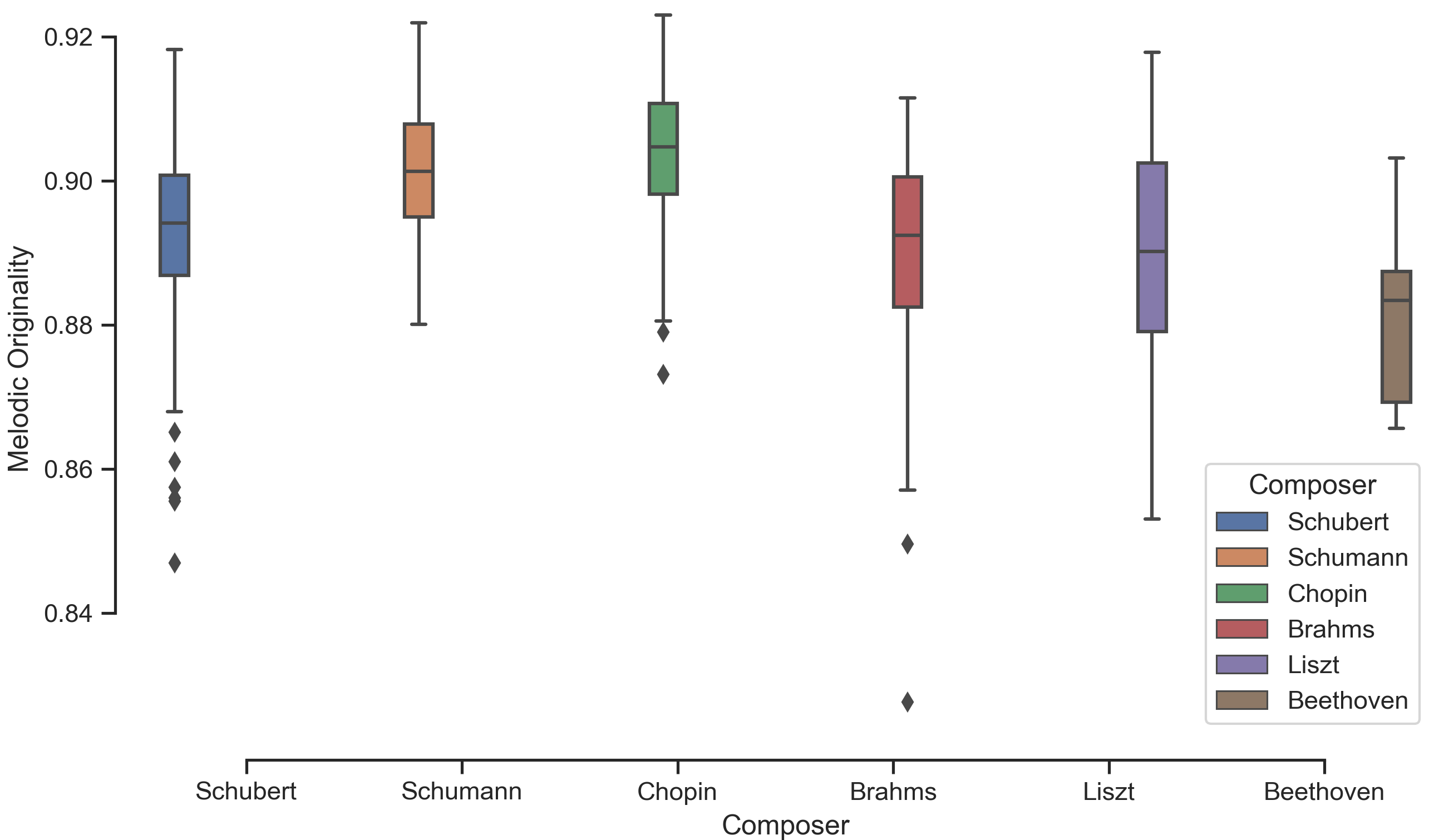}
  \caption{Box plot for melodic originality between all six composers.}
  \label{fig:box}
\end{figure}

In \autoref{fig:scatter} the scatter plot displays melodic originality and popularity with each different color corresponding to a different composer. From this plot, significant differences in the melodic originality between composers are shown.
This is more clearly represented in \autoref{fig:box}. It is clear from this that the novel method for calculating melodic originality is successful in quantifying the stylistic differences between composers. This can be shown statistically with a Two-sided T-Test for the difference in mean melodic originality between all composers analyzed.

\begin{table}[ht]
\centering
\caption{T-Test Results for Difference in Mean Melodic Originality between Composers}
\label{tab:tests}
\begin{tabular}{rll}
  \hline
 & Test & Results \\ 
  \hline
1 & Schumann and Liszt: & t=3.9945, p=0.0001 \\
\hline
2 & Schumann and Beethoven: & t=7.2988, p=0.0 \\
\hline
3 & Schumann and Schubert: & t=4.7214, p=0.0 \\
\hline
4 & Schumann and Chopin: & t=-0.9682, p=0.3343 \\
\hline
5 & Schumann and Brahms: & t=5.348, p=0.0 \\
\hline
6 & Liszt and Beethoven: & t=2.2753, p=0.0283 \\
\hline
7 & Liszt and Schubert: & t=-0.6302, p=0.5296 \\
\hline
8 & Liszt and Chopin: & t=-5.414, p=0.0 \\
\hline
9 & Liszt and Brahms: & t=0.5253, p=0.6003 \\
\hline
10 & Beethoven and Schubert: & t=-3.4409, p=0.0008 \\
\hline
11 & Beethoven and Chopin: & t=-8.5685, p=0.0 \\
\hline
12 & Beethoven and Brahms: & t=-1.8819, p=0.0625 \\
\hline
13 & Schubert and Chopin: & t=-7.2988, p=0.0 \\
\hline
14 & Schubert and Brahms: & t=1.8302, p=0.0687 \\
\hline
15 & Chopin and Brahms: & t=8.0819, p=0.0 \\
\hline

\end{tabular}
\end{table}

In \autoref{tab:tests} it is shown that 10 out of the 15 tests were significant with $\alpha$ = 0.05. This shows that there is a significant statistical difference between mean melodic originality for a majority of the composers tested. This provides clear evidence that the novel method for calculating melodic originality in this paper is effective at detecting different styles.

The most original pieces were also recorded for this research and are shown in \autoref{tab:ranked}.

\begin{table}[ht]
\centering
\caption{Highest Scoring Pieces for Melodic Originality}
\label{tab:ranked}
\begin{tabular}{rlll}
  \hline
 & Title & Composer & Originality \\ 
  \hline
1 & Mazurka, Op.50, No.2 & Chopin & 0.9229 \\
\hline
2 & 7 Klavierstücke, Op.126, No. 1 & Schumann & 0.9219 \\
\hline
3 & Mazurka, Op.17, No.3 & Chopin &  0.9217\\
\hline
4 & 4 Fugues, Op.72, No. 3 & Schumann & 0.9214 \\
\hline
5 & Valse Op.64 No. 1 & Chopin & 0.9189\\
\hline

\end{tabular}
\end{table}

Chopin had the highest mean melodic originality, as well as three pieces in the top five. Schumann took the other two spots. Overall, Schumann and Chopin had the highest mean melodic originality scores, which is why \autoref{tab:tests} shows no statistical difference between their mean melodic originality.

\subsection{Figures}


\section{Conclusion}

From this data, we can conclude that there is no correlation between melodic originality and thematic fame in the 428 Romantic period pieces that were analyzed. This contrasts Simonton’s finding that melodic originality and thematic fame have a curvilinear relationship. These different conclusions represent our different methods of calculating melodic originality. While Simonton only focused on the first 6 notes, the novel method in this paper accounts for all notes in the piece. This provides much more data and results that indicate no correlation between popularity and melodic originality. It is possible that the small sample size of n=428 compared to Simonton’s n=15,618 could be the cause of a difference in conclusions. This novel method for calculating melodic originality should be further explored and tested to confirm its viability in relation to popularity. Despite these findings, it does appear that this method for calculating melodic originality is effective in quantifying the stylistic differences between composers. This can be concluded from the results that most composers analyzed have statistically significant differences in mean melodic originality.

\begin{table*}
  \centering
  \caption{Total Note-Transition Counts}
  \label{tab:totals}
    \begin{tabular}{|l|l|l|l|l|l|l|l|l|l|l|l|l|}
    \hline
        & C       & C\#     & D       & D\#     & E       & F       & F\#     & G       & G\#     & A       & A\#     & B       \\ \hline
    C   & 48706 & 9549  & 15741 & 21873 & 17435 & 19708 & 5615  & 16120 & 12433 & 15017 & 16698 & 12510 \\ \hline
    C\# & 10160 & 31841 & 12319 & 10850 & 17553 & 12916 & 12180 & 5707  & 12410 & 10115 & 12691 & 11700 \\ \hline
    D   & 17350 & 11410 & 51685 & 11681 & 16920 & 23252 & 14940 & 20143 & 6316  & 15870 & 15685 & 14720 \\ \hline
    D\# & 16645 & 12437 & 11874 & 38206 & 9379  & 13576 & 14856 & 15827 & 16686 & 7023  & 15041 & 8077  \\ \hline
    E   & 12539 & 13427 & 17953 & 8298  & 38359 & 12902 & 12134 & 21184 & 13355 & 20445 & 6111  & 13862 \\ \hline
    F   & 15191 & 10801 & 17276 & 15191 & 14300 & 46562 & 8249  & 15828 & 17335 & 16654 & 19602 & 6221  \\ \hline
    F\# & 5734  & 10903 & 10477 & 12231 & 13489 & 8066  & 30101 & 10727 & 10150 & 18456 & 9809  & 13346 \\ \hline
    G   & 20957 & 6371  & 15137 & 13084 & 15632 & 17591 & 10659 & 48778 & 9961  & 15464 & 19816 & 14701 \\ \hline
    G\# & 16340 & 13814 & 6615  & 14083 & 10312 & 13832 & 11161 & 10080 & 32128 & 11051 & 10538 & 14397 \\ \hline
    A   & 20344 & 13638 & 21398 & 7166  & 15723 & 12624 & 14360 & 16294 & 10742 & 49846 & 12608 & 14166 \\ \hline
    A\# & 14466 & 14997 & 20376 & 16563 & 6073  & 15138 & 8586  & 16744 & 12694 & 12486 & 48808 & 7463  \\ \hline
    B   & 12741 & 9849  & 19341 & 10338 & 17018 & 6227  & 10925 & 10433 & 11288 & 15326 & 6886  & 31510 \\ \hline
    \end{tabular}
\end{table*}

\begin{table*}
  \centering
  \caption{Stochastic Two-Note Transtion Probability Matrix}
  \label{tab:prob}
\begin{tabular}{|l|l|l|l|l|l|l|l|l|l|l|l|l|}
\hline
    & C      & C\#    & D      & D\#    & E      & F      & F\#    & G      & G\#    & A      & A\#    & B      \\ \hline
C   & 0.2304 & 0.0452 & 0.0745 & 0.1035 & 0.0825 & 0.0932 & 0.0266 & 0.0763 & 0.0588 & 0.071  & 0.079  & 0.0592 \\ \hline
C\# & 0.0633 & 0.1985 & 0.0768 & 0.0676 & 0.1094 & 0.0805 & 0.0759 & 0.0356 & 0.0773 & 0.063  & 0.0791 & 0.0729 \\ \hline
D   & 0.0789 & 0.0519 & 0.235  & 0.0531 & 0.0769 & 0.1057 & 0.0679 & 0.0916 & 0.0287 & 0.0721 & 0.0713 & 0.0669 \\ \hline
D\# & 0.0927 & 0.0692 & 0.0661 & 0.2127 & 0.0522 & 0.0756 & 0.0827 & 0.0881 & 0.0929 & 0.0391 & 0.0837 & 0.045  \\ \hline
E   & 0.0658 & 0.0705 & 0.0942 & 0.0435 & 0.2013 & 0.0677 & 0.0637 & 0.1112 & 0.0701 & 0.1073 & 0.0321 & 0.0727 \\ \hline
F   & 0.0748 & 0.0532 & 0.085  & 0.0748 & 0.0704 & 0.2291 & 0.0406 & 0.0779 & 0.0853 & 0.082  & 0.0965 & 0.0306 \\ \hline
F\# & 0.0374 & 0.071  & 0.0683 & 0.0797 & 0.0879 & 0.0526 & 0.1961 & 0.0699 & 0.0661 & 0.1202 & 0.0639 & 0.087  \\ \hline
G   & 0.1007 & 0.0306 & 0.0727 & 0.0629 & 0.0751 & 0.0845 & 0.0512 & 0.2343 & 0.0479 & 0.0743 & 0.0952 & 0.0706 \\ \hline
G\# & 0.0994 & 0.0841 & 0.0402 & 0.0857 & 0.0627 & 0.0842 & 0.0679 & 0.0613 & 0.1955 & 0.0672 & 0.0641 & 0.0876 \\ \hline
A   & 0.0974 & 0.0653 & 0.1024 & 0.0343 & 0.0753 & 0.0604 & 0.0687 & 0.078  & 0.0514 & 0.2386 & 0.0604 & 0.0678 \\ \hline
A\# & 0.0744 & 0.0771 & 0.1048 & 0.0852 & 0.0312 & 0.0779 & 0.0442 & 0.0861 & 0.0653 & 0.0642 & 0.2511 & 0.0384 \\ \hline
B   & 0.0787 & 0.0608 & 0.1195 & 0.0639 & 0.1051 & 0.0385 & 0.0675 & 0.0644 & 0.0697 & 0.0947 & 0.0425 & 0.1946 \\ \hline
\end{tabular}
\end{table*}

\begin{table*}
  \centering
  \caption{OLS Regression Results}
  Ordinary Least Squares Regression
  \label{tab:OLS}
\begin{center}
\begin{tabular}{lclc}
\toprule
\textbf{Dep. Variable:}    & Melodic Originality & \textbf{  R-squared (uncentered):}      &     0.523   \\
\textbf{Model:}            &         OLS         & \textbf{  Adj. R-squared (uncentered):} &     0.522   \\
\textbf{Method:}           &    Least Squares    & \textbf{  F-statistic:       }          &     467.7   \\
\textbf{Date:}             &   Thu, 20 May 2021  & \textbf{  Prob (F-statistic):}          &  1.38e-70   \\
\textbf{Time:}             &       11:38:07      & \textbf{  Log-Likelihood:    }          &   -401.77   \\
\textbf{No. Observations:} &           428       & \textbf{  AIC:               }          &     805.5   \\
\textbf{Df Residuals:}     &           427       & \textbf{  BIC:               }          &     809.6   \\
\textbf{Df Model:}         &             1       & \textbf{                     }          &             \\
\bottomrule
\end{tabular}
\begin{tabular}{lcccccc}
                    & \textbf{coef} & \textbf{std err} & \textbf{t} & \textbf{P$> |$t$|$} & \textbf{[0.025} & \textbf{0.975]}  \\
\midrule
\textbf{Popularity} &    1.246e-06  &     5.76e-08     &    21.626  &         0.000        &     1.13e-06    &     1.36e-06     \\
\bottomrule
\end{tabular}
\begin{tabular}{lclc}
\textbf{Omnibus:}       & 102.896 & \textbf{  Durbin-Watson:     } &    0.310  \\
\textbf{Prob(Omnibus):} &   0.000 & \textbf{  Jarque-Bera (JB):  } &  225.749  \\
\textbf{Skew:}          &  -1.244 & \textbf{  Prob(JB):          } & 9.53e-50  \\
\textbf{Kurtosis:}      &   5.543 & \textbf{  Cond. No.          } &     1.00  \\
\bottomrule
\end{tabular}
\end{center}

\textbf{Notes:}
[1] R² is computed without centering (uncentered) since the model does not contain a constant. \newline
[2] Standard Errors assume that the covariance matrix of the errors is correctly specified.
 
\end{table*}

\clearpage
\bibliographystyle{ACM-Reference-Format}
\bibliography{main}

\end{document}